\documentclass[conference, final, a4paper]{IEEEtran}
\usepackage[nolist]{acronym}
\usepackage{tikz}
\usetikzlibrary{dsp}
\usetikzlibrary{decorations.pathreplacing,calligraphy,calc}
\usetikzlibrary{shapes,arrows,fadings,chains}
\usetikzlibrary{spy}

\usepackage{pgfplots}
\usepackage{pgffor}
\usepgfplotslibrary{groupplots}
\usepackage{graphicx}
\usepackage[font=footnotesize]{caption}
\usepackage{subcaption}
\usepackage{float}
\pgfplotsset{compat=newest}
\usepackage[normalem]{ulem}
\usepackage{amsmath, amsbsy, amssymb}
\usepackage{nicefrac}
\usepackage{tikzscale}
\usepackage{color}
\usepackage{epigraph}
\usepackage{circuitikz}
\usepackage{multirow}
\usepackage{booktabs}
\usepackage[ruled]{algorithm2e}

\definecolor{darkgreen}{rgb}{0.125,0.5,0.169}

\usepackage{marvosym}
\usepackage{bbm}
\usepackage{mathtools}

\tikzset{>=latex}

\renewcommand{\vec}[1]{\mathbf{#1}}

\newcommand{\cv}{\vec{c}}

\newcommand{\nv}{\vec{n}}

\newcommand{\uv}{\vec{u}}

\newcommand{\xv}{\vec{x}}
\newcommand{\yv}{\vec{y}}

\newcommand{\Lm}{\vec{L}}

\newcommand{\RR}{\mathbb{R}}

\begin{acronym}
 \acro{CSI}{channel state information}
 \acro{UE}{user equipment}
 \acro{UL}{uplink}
 \acro{BS}{basestation}
 \acro{TDD}{time division duplex}
 \acro{FDD}{frequency division duplex}
 \acro{ECC}{error-correcting code}
 \acro{MLD}{maximum likelihood decoding}
 \acro{HDD}{hard decision decoding}
 \acro{IF}{intermediate frequency}
 \acro{RF}{radio frequency}
 \acro{SDD}{soft decision decoding}
 \acro{NND}{neural network decoding}
 \acro{CNN}{convolutional neural network}
 \acro{ML}{maximum likelihood}
 \acro{GPU}{graphical processing unit}
 \acro{BP}{belief propagation}
 \acro{LTE}{Long Term Evolution}
 \acro{BER}{bit error rate}
 \acro{DER}{detection error rate}
 \acro{SNR}{signal-to-noise-ratio}
 \acro{ReLU}{rectified linear unit}
 \acro{BPSK}{binary phase shift keying}
 \acro{QPSK}{quadrature phase shift keying}
 \acro{AWGN}{additive white Gaussian noise}
 \acro{MSE}{mean squared error}
 \acro{LLR}{log-likelihood ratio}
 \acro{MAP}{maximum a posteriori}
 \acro{NVE}{normalized validation error}
 \acro{BCE}{binary cross-entropy}
 \acro{CE}{cross-entropy}
 \acro{BLER}{block error rate}
 \acro{SQR}{signal-to-quantisation-noise-ratio}
 \acro{MIMO}{multiple-input multiple-output}
 \acro{OFDM}{orthogonal frequency division multiplex}
 \acro{RF}{radio frequency}
 \acro{LOS}{line of sight}
 \acro{NLoS}{non-line of sight}
 \acro{NMSE}{normalized mean squared error}
 \acro{CFO}{carrier frequency offset}
 \acro{SFO}{sampling frequency offset}
 \acro{IPS}{indoor positioning system}
 \acro{TRIPS}{time-reversal IPS}
 \acro{RSSI}{received signal strength indicator}
 \acro{MIMO}{multiple-input multiple-output}
 \acro{ENoB}{effective number of bits}
 \acro{AGC}{automated gain control}
 \acro{ADC}{analog to digital converter}
 \acro{ADCs}{analog to digital converters}
 \acro{FB}{front bandpass}
 \acro{FPGA}{field programmable gate array}
 \acro{JSDM}{Joint Spatial Division and Multiplexing}
 \acro{NN}{neural network}
 \acro{IF}{intermediate frequency}
 \acro{LoS}{line-of-sight}
 \acro{NLoS}{non-line-of-sight}
 \acro{DSP}{digital signal processing}
 \acro{AFE}{analog front end}
 \acro{SQNR}{signal-to-quantisation-noise-ratio}
 \acro{SINR}{signal-to-interference-noise-ratio}
 \acro{ENoB}{effective number of bits}
 \acro{AGC}{automated gain control}
 \acro{PCB}{printed circuit board}
 \acro{EVM}{error vector mangnitude}
 \acro{CDF}{cumulative distribution function}
 \acro{MRC}{maximum ratio combining}
 \acro{MRP}{maximum ratio precoding}
 \acro{MRT}{maximum ratio transmission}
 \acro{DeepL}{deep-learning}
 \acro{DL}{deep learning}
 \acro{SISO}{single-input single-output}
 \acro{SGD}{stochastic gradient descent}
 \acro{CP}{cyclic prefix}
 \acro{MISO}{Multiple Input Single Output}
 \acro{LMMSE}{linear minimum mean square error}
 \acro{ZF}{zero forcing}
 \acro{USRP}{universal software radio peripheral}
 \acro{RNN}{recurrent neural network}
 \acro{GRU}{gated recurrent unit}
 \acro{LSTM}{long short-term memory}
 \acro{NTM}{neural turing machine}
 \acro{DNC}{differentiable neural computer}
 \acro{TCN}{temporal convolutional network}
 \acro{FCL}{fully connected layer}
 \acro{MANN}{memory augmented neural network}
 \acro{RNN}{recurrent neural network}
 \acro{DNN}{dense neural network}
 \acro{FIR}{finite impulse response}
 \acro{BPTT}{back-propagation through time}
 \acro{GAN}{generative adversarial network}
 \acro{ELU}{exponential linear unit}
 \acro{tanh}{hyperbolic tangent}
 \acro{BICM}{bit-interleaved coded modulation}
 \acro{OTA}{over-the-air}
 \acro{IM}{intensity modulation}
 \acro{DD}{direct detection}
 \acro{RL}{reinforcement learning}
 \acro{SDR}{software-defined radio}
 \acro{WGAN}{Wasserstein generative adversarial network}
 \acro{BMD}{bit-metric decoding}
 \acro{BMI}{bit-wise mutual information}
 \acro{LDPC}{low-density parity-check}
 \acro{IDD}{iterative demapping and decoding}
 \acro{IEDD}{iterative equalization, demapping and decoding}
 \acro{JSD}{Jensen-Shannon divergence}
 \acro{MMSE}{minimum mean square error}
 \acro{FFT}{fast Fourier transform}
 \acro{IFFT}{inverse fast Fourier transform}
 \acro{QAM}{quadrature amplitude modulation}
 \acro{EMD}{earth mover's distance}
 \acro{TDL}{tapped delay line}
 \acro{KL}{Kullback-Leibler}
 \acro{PRACH}{physical random access channel}
 \acro{URLLC}{ultra-reliable low-latency communication}
 \acro{ANOMA}{asynchronous non-orthogonal multiple access}
 \acro{FEC}{forward error correction}
 \acro{NOMA}{non-orthogonal medium access}
 \acro{MTC}{machine-type communications}
 \acro{mMTC}{massive machine-type communications}
 \acro{MCS}{modulation and coding scheme}
 \acro{PAPR}{peak-to-average power ratio}
 \acro{MAC}{medium access control}
 \acro{STO}{sampling time offset}
 \acro{STE}{straight-through estimator}
 \acro{PHY}{physical}
 \acro{CCE}{categorical cross-entropy}
 \acro{IoT}{Internet of Things}
 \acro{CCDF}{complementary cumulative distribution function}
 \acro{CRC}{cyclic redundancy check}
 \acro{ACLR}{adjacent channel leakage ratio}
\acro{MD}{missed detection}
 \acro{FA}{false alarm}
 \acro{FAR}{false alarm rate}
 \acro{BCJR}{Bahl-Cocke-Jelinek-Raviv}
 \acro{SCNC}{serially concatenated neural code}
 \acro{TGP}{training with Gaussian priors}
 \acro{DCCNN}{densely connected convolutional neural network}
 \acro{EXIT}{extrinsic information transfer}
\end{acronym}

\definecolor{mittelblau}{RGB}{0, 126, 198}
\definecolor{violettblau}{cmyk}{0.9, 0.6, 0, 0}
\definecolor{rot}{RGB}{238, 28 35}
\definecolor{apfelgruen}{RGB}{140, 198, 62}
\definecolor{gelb}{RGB}{1, 221, 0}
\definecolor{orange}{RGB}{244, 111, 33}
\definecolor{pink}{RGB}{237, 0, 140}
\definecolor{lila}{RGB}{128, 10, 145}
\definecolor{hellgrau}{RGB}{224, 224, 224}
\definecolor{mittelgrau}{RGB}{128, 128, 128}
\definecolor{dunkelgrau}{RGB}{80,80,80}
\definecolor{anthrazit}{RGB}{19, 31, 31}

\pgfplotscreateplotcyclelist{corporate colours markers}{%
rot, every mark/.append style={fill=.!80!rot},mark=*\\%
mittelblau, every mark/.append style={fill=.!80!mittelblau},mark=square*\\%
apfelgruen, every mark/.append style={fill=.!80!apfelgruen},mark=triangle*\\%
orange, mark=star\\%
pink, every mark/.append style={fill=.!80!pink},mark=diamond*\\%
violettblau, every mark/.append style={fill=.!80!violettblau},mark=otimes*\\%
lila, mark=|\\%
gelb, every mark/.append style={fill=.!80!gelb},mark=pentagon*\\%
hellgrau, mark=text,text mark=p\\%
anthrazit, mark=text,text mark=a\\%
}

\pgfplotscreateplotcyclelist{corporate colours markers double}{%
rot, every mark/.append style={fill=.!80!rot},mark=*\\%
rot, dashed, every mark/.append style={fill=.!80!rot,solid},mark=*\\%
mittelblau, every mark/.append style={fill=.!80!mittelblau},mark=square*\\%
mittelblau, dashed, every mark/.append style={fill=.!80!mittelblau,solid},mark=square*\\%
apfelgruen, every mark/.append style={fill=.!80!apfelgruen},mark=triangle*\\%
apfelgruen, dashed, every mark/.append style={fill=.!80!apfelgruen,solid},mark=triangle*\\%
orange, mark=star\\%
orange, dashed, mark=star\\%
pink, every mark/.append style={fill=.!80!pink},mark=diamond*\\%
pink, dashed, every mark/.append style={fill=.!80!pink,solid},mark=diamond*\\%
violettblau, every mark/.append style={fill=.!80!violettblau},mark=otimes*\\%
violettblau, dashed, every mark/.append style={fill=.!80!violettblau,solid},mark=otimes*\\%
lila, mark=|\\%
lila, dashed, mark=|\\%
gelb, every mark/.append style={fill=.!80!gelb},mark=pentagon*\\%
gelb, dashed, every mark/.append style={fill=.!80!gelb,solid},mark=pentagon*\\%
}

\pgfplotsset{
  colormap/magma/.style={%
    /pgfplots/colormap={magma}{%
      rgb=(0.001462, 0.000466, 0.013866)
      rgb=(0.035520, 0.028397, 0.125209)
      rgb=(0.102815, 0.063010, 0.257854)
      rgb=(0.191460, 0.064818, 0.396152)
      rgb=(0.291366, 0.064553, 0.475462)
      rgb=(0.384299, 0.097855, 0.501002)
      rgb=(0.475780, 0.134577, 0.507921)
      rgb=(0.569172, 0.167454, 0.504105)
      rgb=(0.664915, 0.198075, 0.488836)
      rgb=(0.761077, 0.231214, 0.460162)
      rgb=(0.852126, 0.276106, 0.418573)
      rgb=(0.925937, 0.346844, 0.374959)
      rgb=(0.969680, 0.446936, 0.360311)
      rgb=(0.989363, 0.557873, 0.391671)
      rgb=(0.996580, 0.668256, 0.456192)
      rgb=(0.996727, 0.776795, 0.541039)
      rgb=(0.992440, 0.884330, 0.640099)
      rgb=(0.987053, 0.991438, 0.749504)
    },
  },
  colormap/inferno/.style={%
    /pgfplots/colormap={inferno}{%
      rgb=(0.001462, 0.000466, 0.013866)
      rgb=(0.037668, 0.025921, 0.132232)
      rgb=(0.116656, 0.047574, 0.272321)
      rgb=(0.217949, 0.036615, 0.383522)
      rgb=(0.316282, 0.053490, 0.425116)
      rgb=(0.410113, 0.087896, 0.433098)
      rgb=(0.503493, 0.121575, 0.423356)
      rgb=(0.596940, 0.154848, 0.398125)
      rgb=(0.688653, 0.192239, 0.357603)
      rgb=(0.775059, 0.239667, 0.303526)
      rgb=(0.851384, 0.302260, 0.239636)
      rgb=(0.912966, 0.381636, 0.169755)
      rgb=(0.956852, 0.475356, 0.094695)
      rgb=(0.981895, 0.579392, 0.026250)
      rgb=(0.987464, 0.690366, 0.079990)
      rgb=(0.973088, 0.805409, 0.216877)
      rgb=(0.947594, 0.917399, 0.410665)
      rgb=(0.988362, 0.998364, 0.644924)
    },
  },
  colormap/plasma/.style={%
    /pgfplots/colormap={plasma}{%
      rgb=(0.050383, 0.029803, 0.527975)
      rgb=(0.186213, 0.018803, 0.587228)
      rgb=(0.287076, 0.010855, 0.627295)
      rgb=(0.381047, 0.001814, 0.653068)
      rgb=(0.471457, 0.005678, 0.659897)
      rgb=(0.557243, 0.047331, 0.643443)
      rgb=(0.636008, 0.112092, 0.605205)
      rgb=(0.706178, 0.178437, 0.553657)
      rgb=(0.768090, 0.244817, 0.498465)
      rgb=(0.823132, 0.311261, 0.444806)
      rgb=(0.872303, 0.378774, 0.393355)
      rgb=(0.915471, 0.448807, 0.342890)
      rgb=(0.951344, 0.522850, 0.292275)
      rgb=(0.977856, 0.602051, 0.241387)
      rgb=(0.992541, 0.687030, 0.192170)
      rgb=(0.992505, 0.777967, 0.152855)
      rgb=(0.974443, 0.874622, 0.144061)
      rgb=(0.940015, 0.975158, 0.131326)
    },
  },
}

\IEEEoverridecommandlockouts

\begin{document}

\title{Component Training of Turbo Autoencoders}

\author{\IEEEauthorblockN{Jannis Clausius, Marvin Geiselhart and Stephan ten Brink}

\IEEEauthorblockA{
 Institute of Telecommunications, University of Stuttgart, Pfaffenwaldring 47, 70659 Stuttgart, Germany \\
\{clausius,geiselhart,tenbrink\}@inue.uni-stuttgart.de\\
}

\thanks{This work is supported by the German Federal Ministry of Education and Research (BMBF) within the project Open6GHub under grant 16KISK019 and the project FunKI  under grant 16KIS1187.}
}

\maketitle
\begin{abstract}
Isolated \ac{TGP} of the component autoencoders of turbo-autoencoder architectures  enables faster, more consistent training and better generalization to arbitrary decoding iterations than training based on deep unfolding.
We propose fitting the components via \ac{EXIT} charts to a desired behavior which enables scaling to larger message lengths ($k \approx 1000$) while retaining competitive performance.
To the best of our knowledge, this is the first autoencoder that performs close to classical codes in this regime.
Although the \ac{BCE} loss function optimizes the \ac{BER} of the components, the design via \ac{EXIT} charts enables to focus on the \ac{BLER}.
In serially concatenated systems the component-wise \ac{TGP} approach is well known for inner components with a fixed outer binary interface, e.g., a learned inner code or equalizer, with an outer binary error correcting code.
In this paper we extend the component training to structures with an inner and outer autoencoder, where we propose a new 1-bit quantization strategy for the encoder outputs based on the underlying communication problem.
Finally, we discuss the model complexity of the learned components during design time (training) and inference and show that the number of weights in the encoder can be reduced by $99.96\,\%$. 
\end{abstract}

\acresetall

\section{Introduction}
Concatenation of blocks, or modularization, is a crucial concept in engineering that breaks down complex systems into smaller, more manageable components to simplify the design process. 
In communications engineering, iterating between the concatenated blocks of a receiver gives rise to an even more powerful concept: the \textit{turbo principle} \cite{turbo02hagenauer}. Here, two or more serially concatenated blocks iteratively exchange extrinsic information.
Since its introduction in turbo decoders \cite{berrou1993near}, the principle found application in many different scenarios, such as turbo equalization of multipath channels, \ac{MIMO} detection, \ac{BICM} and \ac{LDPC} decoding \cite{turbo02hagenauer}.
These systems show outstanding performance close to the theoretical limits in scenarios closely matching the models assumed in the design, e.g., the channel model.
However, in more complex situations or scenarios with hardware impairments or when no suitable channel models is known, machine learning-aided communication systems that are trained to optimize the end-to-end performance can outperform classical systems \cite{o2017introduction, doerner2022jointautoencoder}. Still, the well-understood \ac{AWGN} scenario serves as a valuable benchmark and can be seen as the ``worst case'' scenario for the learning-based systems.
Like in conventional communications, the turbo principle has been successfully applied in form of the turbo-autoencoder, called TurboAE, to scale deep-learning based transceivers to more practical block lengths \cite{jiang19turboae}.
The TurboAE consists of parallel or serially concatenated \acp{CNN} at the transmitter. At the receiver, information is passed iteratively between another set of two \acp{CNN}. It is typically trained in a deep unfolded fashion, i.e., the iterations of the receiver are unrolled into a deep \ac{NN}, optimally adapting the constituent \acp{NN} to the iterative algorithm. 
However, this comes at the cost of increased training complexity which scales linearly with the number of iterations. 
To overcome this issue, component \ac{TGP} has been proposed in \cite{pretrain20koike, clausius21serialAE}.

In this paper we consider component-wise autoencoder training for the serial TurboAE. %
The main contributions of this paper are as follows:

\begin{itemize}
   \item We apply \ac{EXIT} charts \cite{exit01brink} to analyze and optimize the serial TurboAE for \ac{BLER} or large block lengths by finetuning the inner component to the outer autoencoder \cite{exitmap98brink}. %
   \item A Gaussian prior training framework for the serial TurboAE is proposed and demonstrated to drastically reduce the training complexity.
   \item A new binarization strategy for an encoder output layer (\ac{BPSK} modulation) is proposed based on the underlying communication problem.
   \item The encoder networks are distilled down to just 148 weights by a student-teacher method without performance degradation, enabling practical implementations. %
\end{itemize}

\section{Preliminaries}
\label{sec:sys}

\subsection{Densely Connected Convolutional Layers}
\Acp{DCCNN} \cite{Huang2017DenseNet} allow the training of deeper structures with fewer weights than plain \acp{CNN}.
A \ac{DCCNN} consists of blocks of convolutional layers with an increased amount of connections, and transition layers.
The input to each densely connected convolutional layer is not only the output from the last layer, but a concatenation of all preceding feature maps within a block.
This means the number of feature maps for inputs increases with a growth parameter $F$, which is the number of output feature maps per layer.
Furthermore, each layer consists of a batch normalization, a \ac{ReLU} activation and a convolutional layer with kernel size $K$.
Every densely connected block proceeds a transition layer.
The output of the transition layer is the only input to the next densely connected block.
It interrupts the growth process of the number of feature maps. 
The transition layer is a $1\times1$ convolutional layer that outputs $F_0$ feature maps.

\subsection{Serial Turbo Autoencoder}

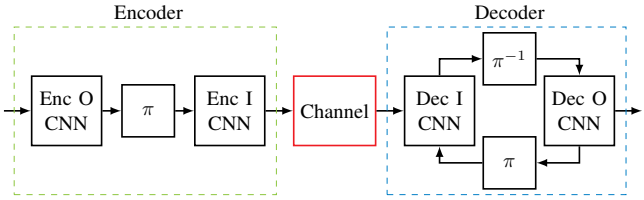
\begin{figure}[t]
	\centering
	\resizebox{\linewidth}{!}{\begin{tikzpicture}
    \tikzstyle{box} =[rectangle,  draw=black, thick, align=flush center,minimum width=1.2cm, minimum height=1.2cm, text centered]; 
    \tikzstyle{intbox} =[rectangle,  draw=black, thick, align=flush center,minimum width=0.9cm, minimum height=0.9cm, text centered]; 
    \tikzstyle{arrow} = [thick,->]
	
	\node [] (start) at (0,0) {};	
	\node [box] (enco) at (1.2,0) {Enc O\\CNN};
	\node [intbox] (inte) at (2.6,0) {$\pi$};
	\node [box] (enci) at (4,0) {Enc I\\CNN};
	\node [box, rot,  text=black] (channel) at (5.8,0) {Channel};
 	\node [box] (deci) at (7.6,0) {Dec I\\CNN};
 	\node [intbox] (int1) at (8.8,0.9) {$\pi^{-1}$};
 	\node [intbox] (int2) at (8.8,-0.9) {$\pi$};
 	\node [box] (deco) at (10,0) {Dec O\\CNN};
    \node [] (end) at (11.2,0) {};
	
	\draw [arrow] (start) -- (enco);
	\draw [arrow] (enco) -- (inte);
	\draw [arrow] (inte) -- (enci);
	\draw [arrow] (enci) -- (channel);
	\draw [arrow] (channel) -- (deci);
	\draw [arrow] (deci) |- (int1);
	\draw [arrow] (int1) -| (deco);
	\draw [arrow] (deco) |- (int2);
	\draw [arrow] (int2) -| (deci);
	\draw [arrow] (deco) -- (end);
	
	\draw[draw=apfelgruen, dashed] (0.3,1.45) rectangle (4.8,-1.45);
	\draw[draw=mittelblau, dashed] (6.7,1.45) rectangle (10.8,-1.45);
	
	\node[] at (2.6,1.7) {Encoder};
	\node[] at (8.8,1.7) {Decoder};

\end{tikzpicture}}
	\vspace{-0.15cm}
	\caption{\footnotesize Structure of the CNN-based serial TurboAE communication system consisting of a real-valued encoder, channel, and an iterative decoder.
	}
	\label{fig:serial_autoencoder}
	\vspace{-0.25cm}
\end{figure}

The serial TurboAE \cite{clausius21serialAE} is based on the structure of serially concatenated codes with an iterative (turbo) decoding architecture.
However, the encoders and decoders are implemented as \acp{CNN}.
The system model is shown in Fig.~\ref{fig:serial_autoencoder}. 
The encoder consists of an outer encoder, an interleaver and an inner encoder. 
The corresponding decoder is based on a \ac{CNN} for decoding the inner code and a \ac{CNN} for decoding the outer code. 
Again, both are connected by an interleaver and a deinterleaver.
For training, the iterations of the decoder are unfolded, yielding a deep neural decoder with intermediate (de-)interleaver layers.
The difference to the straight-forward training with \ac{SGD} is the alternating training schedule for the encoder ($T_\mathrm{TX}$ updates) and the decoder ($T_\mathrm{RX}$ updates). 
For further details, the reader is referred to \cite{jiang19turboae}.

\subsection{Training with Gaussian Priors}

\ac{TGP} \cite{pretrain20koike, clausius21serialAE} is a method to reduce complexity during training for systems with concatenated components in contrast to unfolding the decoder. 
The idea is to train each component individually and isolated with a single decoding iteration. %
Only during inference the components are concatenated as in the desired system.
The alignment of the components can be done with \ac{EXIT} charts.
For training, the a priori information from the other component has to be generated artificially. 
The generation process is well known under two conditions. First, the information, in form of \acp{LLR}, is Gaussian distributed. Second, the variable for which we generate the a priori \acp{LLR} consists of bits. 
The a priori \ac{LLR} distribution $L_u^\mathrm{A}$ for a bit $u$ and a set a priori information $I_\mathrm{A}$ is then given by   
\begin{align}
\label{eq:generate_prior}
L_u^\mathrm{A} &\sim \mathcal{N}\left((2u-1)\cdot \mu(I_\mathrm{A}), 2\mu(I_\mathrm{A})\right)
\end{align}
where $\mu(I_\mathrm{A})\approx\frac{1}{2}\left(-\frac{1}{H_1}  \log_2 \left( 1-I_\mathrm{A}^{\frac{1}{H_3}}  \right)  \right)^\frac{1}{H_2}$ with constants $H_1=0.3073$, $H_2=0.8935$ and $H_3=1.1064$ \cite{conv05brann}.

\subsection{EXIT Charts}
EXIT charts \cite{exit01brink} are a design tool for iterative, component based systems.
The key idea is to calculate the input/output behavior of each component and predict the decoding trajectory of the concatenated system.
The behavior is displayed by the produced extrinsic information from a certain a priori information. 
The a priori \acp{LLR} can be generated according to (\ref{eq:generate_prior}) and the information between \acp{LLR} $\Lm$ and bits $\uv$ can be approximated \cite{turbo02hagenauer} by
\begin{equation}
    I(\Lm, \uv) \approx 1- \sum_i \log_2 \left( 1+\exp(-(2u_i-1) \cdot L_i) \right).
\end{equation}
The chart shows these two curves where one is flipped along the first bisector.
This means that the x-axis shows the a priori information of one component and the extrinsic information from the other component and vice versa for the y-axis.
The decoding trajectory can be estimated by iterating ``ping-pong''-wise between the two curves. 
The intersection of the curves indicates the maximum reachable mutual  information by iterating between the components. 
However, the estimated trajectory is only accurate if the exchanged information is uncorrelated between the components. %
This does not hold for short block lengths, as cycles in the decoding graph lead to correlated information exchange. %
A more appropriate design tool for short block lengths are scattered EXIT charts \cite{ebada2018scatteredexit}.
Here, the chart consists of many trajectories displayed as a scatter plot.

\section{Component Training for Serial Architectures}
\label{sec:open_loop_serial}

\subsection{Identifying Component Interfaces}

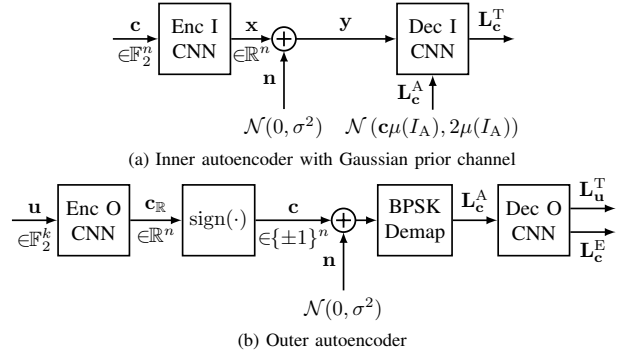
\begin{figure}
     \centering
     \begin{subfigure}[b]{\columnwidth}
         \centering
         \resizebox{0.65\linewidth}{!}{\begin{tikzpicture}
    \tikzstyle{box} =[rectangle,  draw=black, thick, align=flush center,minimum width=1.2cm, minimum height=1.2cm, text centered]; 
    \tikzstyle{arrow} = [thick,->]
	
	\node [] (start) at (0.5,0) {};	
	\node [box] (enc) at (2,0) {Enc I\\CNN};
	\node [dspadder] (channel) at (3.5,0) {};
 	\node [] (n) at (3.5,-1.5) {$\mathcal{N}(0,\sigma^2)$};	
 	\node [] (la) at (6,-1.5) {$\mathcal{N}\left(\cv  \mu(I_\mathrm{A}), 2\mu(I_\mathrm{A})\right)$};
	\node [box] (dec) at (6,0) {Dec I\\ CNN};
    \node [] (end) at (7.5,0) {};
	
	\draw [arrow] (start)-- node[midway, above]{$\cv$}node[below]{$\in\!\!\mathbb{F}_2^{n}$} (enc);
	\draw [arrow] (enc)-- node[midway, above]{$\xv$}node[below]{$\in\!\!\RR^{n}$} (channel);
	\draw [arrow] (channel)-- node[midway, above]{$\yv$} (dec);
	\draw [arrow] (dec)-- node[midway, above]{$\Lm^\mathrm{T}_\cv$} (end);
	\draw [arrow] (la)-- node[midway, left]{$\Lm^\mathrm{A}_\cv$} (dec);
	\draw [arrow] (n) -- node[midway, left]{$\nv$} (channel);
	
\end{tikzpicture}}
         \vspace{-0.15cm}
         \caption{Inner autoencoder with Gaussian prior channel}
         \label{fig:seperate_inner}
     \end{subfigure}
     \hfill
         \begin{subfigure}[b]{\columnwidth}
         \centering
         \resizebox{0.95\linewidth}{!}{\begin{tikzpicture}
    \tikzstyle{box} =[rectangle,  draw=black, thick, align=flush center,minimum width=1.2cm, minimum height=1.2cm, text centered]; 
    \tikzstyle{arrow} = [thick,->]
	
	\node [] (start) at (0.1,0) {};	
	\node [box] (enc) at (1.6,0) {Enc O\\CNN};
	\node [box] (bin) at (3.7,0) {$\operatorname{sign}(\cdot)$};
	\node [dspadder] (channel) at (5.8,0) {};
 	\node [] (n) at (5.8,-1.5) {$\mathcal{N}(0,\sigma^2)$};	
 	\node [box] (demap) at (7,0) {BPSK\\Demap};
 	\node [box] (dec) at (9,0) {Dec O\\CNN};
    \node [] (end1) at (10.5,0.2) {};
    \node [] (end2) at (10.5,-0.2) {};
	
	\draw [arrow] (start) -- node[midway, above]{$\uv$}node[below]{$\in\!\!\mathbb{F}_2^{k}$}  (enc);
	\draw [arrow] (enc) -- node[midway, above]{$\cv_\mathbb{R}$}node[below]{$\in\!\!\RR^{n}$} (bin);
	\draw [arrow] (bin) -- node[midway, above]{$\cv$}node[below]{$\in\!\!\{\pm1\}^{n}$} (channel);
	\draw [arrow] (channel) -- (demap);
	\draw [arrow] (demap) -- node[midway, above]{$\Lm^\mathrm{A}_\cv$} (dec);
	\draw [arrow] (dec.east |- end1) -- node[midway, above]{$\Lm^\mathrm{T}_\uv$} (end1);
	\draw [arrow] (dec.east |- end2) -- node[midway, below]{$\Lm^\mathrm{E}_\cv$} (end2);
	\draw [arrow] (n) -- node[midway, left]{$\nv$} (channel);
	
\end{tikzpicture}}
         \vspace{-0.15cm}
         \caption{Outer autoencoder}
         \label{fig:seperate_outer}
     \end{subfigure}
     
    \vspace{-0.15cm}
    \caption{Block diagrams of the (non-iterative) autoencoders. }
    \vspace{-0.25cm}
    \label{fig:seperate_training}
\end{figure}

The serial TurboAE can be interpreted as an inner autoencoder, and an outer autoencoder with \ac{BPSK} modulation and an virtual channel in between.
This virtual channel consists of the inner autoencoder and the actual channel, but can be abstracted to an \ac{AWGN} channel with an certain \ac{SNR}.
The concept is shown in Fig.~\ref{fig:seperate_training}.
The goal of the component training is to optimize each autoencoder isolated. 
Thus, we first identify the components and their interfaces.
For a serial TurboAE with rate $R=\frac{k}{n}= R_\mathrm{O}\cdot R_\mathrm{I}=\nicefrac{1}{2}$ we choose the outer and inner autoencoder to have rate $R_\mathrm{O}=\nicefrac{1}{2}$ and $R_\mathrm{I}=1$, respectively.  
\cite{ashikhmin04extrinsic} showed that this allocation of the rate is optimal for asymptotic block lengths and \cite{clausius2023neuralcodeclassicaldecoder} observed it empirically.
The inner autoencoder is shown in Fig.~\ref{fig:seperate_inner}.
The input to the inner decoder are the real-valued channel observations $\yv$ and the available a priori information in form of \acp{LLR} $\Lm^\mathrm{A}_\cv$ from the (artificial) outer decoder.
The output are the extrinsic \acp{LLR} $\Lm^\mathrm{E}_\cv = \Lm^\mathrm{T}_\cv - \Lm^\mathrm{A}_\cv$, where the total \acp{LLR} $\Lm^\mathrm{T}_\cv$ are the direct output of the \ac{NN}.
As loss, we choose a \ac{BCE} loss between $\cv$ and $\Lm^\mathrm{T}_\cv$.
To generate artificial a priori \acp{LLR} for the inner autoencoder, the outer autoencoder must use \ac{BPSK} symbols.
The outer autoencoder is shown in Fig.~\ref{fig:seperate_outer}.
The input to the outer decoder are the a priori \acp{LLR} of the coded bits $\Lm^\mathrm{A}_\cv$. 
The outputs are not only the uncoded bit estimates $\Lm^\mathrm{T}_\uv$, but also the refined extrinsic \acp{LLR} $\Lm^\mathrm{E}_\cv = \Lm^\mathrm{T}_\cv - \Lm^\mathrm{A}_\cv$, where $\Lm^\mathrm{T}_\cv$ is the direct output of the \ac{NN}.
The outputs can be trained with a \ac{BCE} loss between $\uv$ and $\Lm^\mathrm{T}_\uv$, and $\cv$ and $\Lm^\mathrm{T}_\cv$, respectively. 
The losses are added without a weighting factor, as they are in the same order of magnitude.

\subsection{Autoencoder with Binary Modulation }
\label{sec:bpsk_ae}
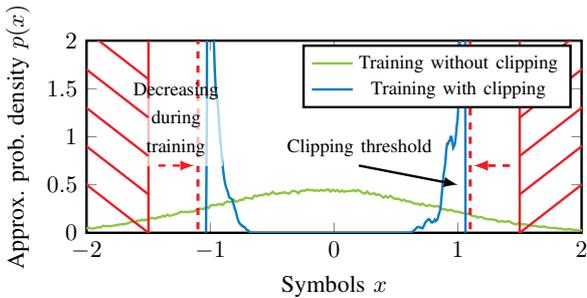
\begin{figure}[t]
	\centering
    \resizebox{0.9\columnwidth}{!}{\begin{tikzpicture}

\begin{axis}[
xmin=-2,
xmax=2,
ymin=0,
ymax=2,
xlabel={Symbols $x$},
ylabel={Approx. prob. density $p(x)$},
width=\linewidth,
height=0.5\linewidth,
line width=1.0pt,
]

\addplot+ [apfelgruen,mark=none] 
table[x=bin ,y=density,col sep=comma]{./tikz/results/training/hist_no_clip.txt};
\addlegendentry{\footnotesize Training without clipping}

\addplot+ [mittelblau,mark=none] 
table[x=bin ,y=smoothed,col sep=comma]{./tikz/results/training/hist_clip105_with_smooth.txt};
\addlegendentry{\footnotesize Training with clipping}

\draw[-latex, color=black] (axis cs:0.2,  0.7) -- node[above,pos=0.01,black, text=black]{\footnotesize Clipping threshold}  (axis cs:1.0, 0.5);

\draw[ color=rot] (axis cs:-1.5,  5) -- node[below,pos=0.01,black]{\footnotesize } (axis cs:-1.5, -5);

\draw[ color=rot] (axis cs:1.5,  5) -- node[below,pos=0.01,black]{\footnotesize } (axis cs:1.5, -5);

\draw[dashed, color=rot,line width=1.3pt] (axis cs:-1.1,  5) -- node[below,pos=0.01,black]{\footnotesize } (axis cs:-1.1, -5);

\draw[dashed, color=rot,line width=1.3pt] (axis cs:1.1,  5) -- node[below,pos=0.01,black]{\footnotesize } (axis cs:1.1, -5);

\draw[draw=none,fill=white,fill opacity=0.7] (axis cs:-1.75,  1.4) rectangle (axis cs:-0.9, 0.7);

\draw[-latex, dashed,color=rot] (axis cs:-1.42,  0.7) -- node[above,dashed,pos=0.5,black, align=center]{\footnotesize Decreasing\\ \footnotesize during\\ \footnotesize  training}  (axis cs:-1.13, 0.7);

\draw[-latex,dashed, color=rot] (axis cs:1.42,  0.7) -- node[above,pos=0.5,black, align=center]{}  (axis cs:1.13, 0.7);

\draw[ color=rot] (axis cs:-1.5,  0) -- node[below,pos=0.01,black]{\footnotesize } (axis cs:-2.5, 1);

\draw[ color=rot] (axis cs:-1.5,  0.4) -- node[below,pos=0.01,black]{\footnotesize } (axis cs:-2.5, 1.4);

\draw[ color=rot] (axis cs:-1.5,  0.8) -- node[below,pos=0.01,black]{\footnotesize } (axis cs:-2.5, 1.8);

\draw[ color=rot] (axis cs:-1.5,  1.2) -- node[below,pos=0.01,black]{\footnotesize } (axis cs:-2.5, 2.2);

\draw[ color=rot] (axis cs:-1.5,  1.6) -- node[below,pos=0.01,black]{\footnotesize } (axis cs:-2.5, 2.6);

\draw[-latex,dashed, color=rot] (axis cs:1.42,  0.7) -- node[above,pos=0.5,black, align=center]{}  (axis cs:1.13, 0.7);

\draw[ color=rot] (axis cs:1.5,  0) -- node[below,pos=0.01,black]{\footnotesize } (axis cs:2.5, 1);

\draw[ color=rot] (axis cs:1.5,  0.4) -- node[below,pos=0.01,black]{\footnotesize } (axis cs:2.5, 1.4);

\draw[ color=rot] (axis cs:1.5,  0.8) -- node[below,pos=0.01,black]{\footnotesize } (axis cs:2.5, 1.8);

\draw[ color=rot] (axis cs:1.5,  1.2) -- node[below,pos=0.01,black]{\footnotesize } (axis cs:2.5, 2.2);

\draw[ color=rot] (axis cs:1.5,  1.6) -- node[below,pos=0.01,black]{\footnotesize } (axis cs:2.5, 2.6);

\end{axis}

\end{tikzpicture}}
    \vspace{-0.15cm}
	\caption{\footnotesize Influence of clipping on the learned modulation.}
	\vspace{-0.15cm}
	\label{fig:clipping}
\end{figure}

Forcing a \ac{BPSK} modulation can be seen as a 1 bit quantization of the encoder output layer. The authors of
\cite{jiang19turboae} propose to first train the output layer with real valued outputs and apply binarization with a  \ac{STE} for gradient computation afterwards.
This procedure outperforms training a model from scratch with  binarization and a  \ac{STE} for gradient computation. %
Here, we propose a new quantization strategy based on the underlying communication problem.
Similar to \cite{jiang19turboae}, we first train with a real valued output layer and only start the quantization process once the system is trained. 
For the latter, we propose to clip $\xv$ with a threshold $x_\mathrm{clip}$, which is gradually decreased from $1.5$ to $1.0$ in steps of $0.1$ every $10$ epochs. %
The key idea is that the clipping is applied after normalization of $\xv$, i.e., reducing the energy of the transmission.
Consequently, the encoder tries to compensate for this energy loss by increasing the probability of symbols with larger magnitude and decreasing the probability of symbols with smaller magnitudes. 
Once $x_\mathrm{clip}\approx 1$, the encoder must converge to a BPSK modulation to transmit with the maximum possible energy. 
For inference we use a binarizer.
The advantage of this approach is that no STE is used, thus, no gradient mismatch between forward and backward path during training.
The concept and intuition is visualized in Fig.~\ref{fig:clipping}.

\subsection{Fitting Components via EXIT Charts}

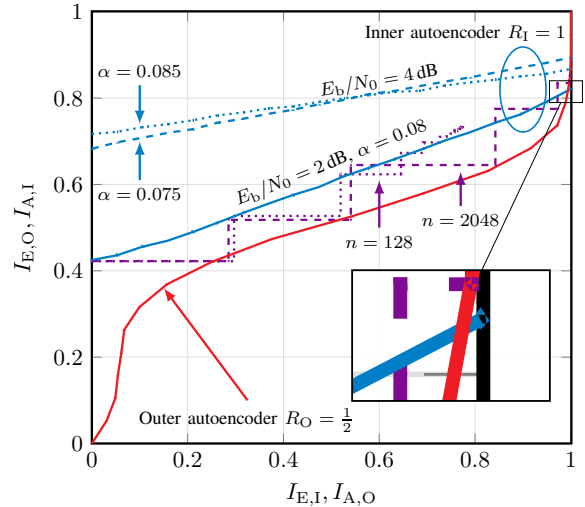
\begin{figure}[t]
	\centering
	\resizebox{0.87\columnwidth}{!}{\begin{tikzpicture}[spy using outlines={rectangle, magnification=6, connect spies}]
	\pgfplotsset{compat=1.5}
	\begin{axis}[
		xmode=normal,
		ymode=normal,
        xlabel={$I_\mathrm{E,I},I_\mathrm{A,O}$},
        ylabel={$I_\mathrm{E,O},I_\mathrm{A,I}$},
		xmin = 0,
		xmax=1.001,
		ymax=1.001,
		ymin=0.0,
		mark size=0.0005pt,
		legend style={at={(axis cs:0,1)},anchor=north west},
		grid=both,
		minor grid style={gray!25},
		major grid style={gray!25},
		width=\columnwidth,
		height=0.925\columnwidth,
		cycle list name=corporate colours markers,
		line width=1pt,
		each nth point=1,
		filter discard warning=false,
		unbounded coords=discard,
		legend cell align={left},
		mark=none
		]

\addplot+ [lila, dotted] 
table[x=x ,y=y,col sep=comma]{./tikz/results/ae_128_precursor_small_snr2_065_precursor_better_decoder_finetune/traj_n128.txt};

\addplot+ [lila, dashed] 
table[x=x ,y=y,col sep=comma]{./tikz/results/ae_128_precursor_small_snr2_065_precursor_better_decoder_finetune/traj_n2048.txt};

\addplot+ [mittelblau, dashed] 
		table[x=mi ,y=e1,col sep=comma]{./tikz/results/ae_128_precursor_17epoch_zpa_075/exit.txt};

\addplot+ [mittelblau, dotted] 
		table[x=mi ,y=e1,col sep=comma]{./tikz/results/ae_128_precursor_17epoch_zpa_085/exit.txt};

\addplot+ [mittelblau] 
table[x=mi ,y=e1,col sep=comma]{./tikz/results/ae_128_precursor_small_snr2_065_precursor_better_decoder_finetune/exit.txt};

\addplot+ [rot] 
table[x=e2 ,y=mi,col sep=comma]{./tikz/results/ae_128_precursor_small_snr2_065_precursor_better_decoder_finetune/exit.txt};

\draw[-latex, line width=1.0pt, color=rot] (axis cs:0.325,  0.1) -- node[below,pos=0.01,black]{\footnotesize Outer autoencoder $R_\mathrm{O}=\frac{1}{2}$} (axis cs:0.15, 0.36);

\draw[-latex, line width=1.0pt, color=lila] (axis cs:0.77,  0.55) -- node[below,pos=0.01,black]{\footnotesize $n=2048$} (axis cs:0.77, 0.63);

\draw[-latex, line width=1.0pt, color=lila] (axis cs:0.6,  0.5) -- node[below,pos=0.01,black]{\footnotesize $n=128$} (axis cs:0.6, 0.61);

\draw[-latex, line width=1.0pt, color=mittelblau] (axis cs:0.1,  0.61) -- node[below,pos=0.01,black]{\footnotesize $\alpha=0.075$} (axis cs:0.1, 0.7);

\draw[-latex, line width=1.0pt, color=mittelblau] (axis cs:0.1,  0.83) -- node[above,pos=0.01,black]{\footnotesize $\alpha=0.085$} (axis cs:0.1, 0.74);

\node[rotate=10] at  (axis cs:0.6, 0.84) {\footnotesize $E_\mathrm{b}/N_0=4\,\text{dB}$};

\node[rotate=22] at  (axis cs:0.5, 0.65) {\footnotesize $E_\mathrm{b}/N_0=2\,\text{dB}$,~$\alpha=0.08$};

\draw[mittelblau, line width=.7pt] (axis cs:0.9, 0.82) ellipse (0.35cm and 0.65cm);
\node at  (axis cs:0.78, 0.95) {\footnotesize Inner autoencoder $R_\mathrm{I}=1$};

\coordinate (spypoint) at (axis cs:0.99
,0.815);
\coordinate (spyviewer) at (axis cs:0.75,0.25);

\end{axis}
	
\spy[width=3cm,height=2cm, fill,fill opacity=0.9] on   (spypoint) in node [fill=white] at (spyviewer);

\end{tikzpicture}	}
	\vspace{-0.15cm}
	\caption{\footnotesize EXIT chart of the learned components.
	}
	\label{fig:exit_chart}
	\vspace{-0.45cm}
\end{figure}

\begin{figure}[t]
	\centering
	\resizebox{0.9\columnwidth}{!}{\begin{tikzpicture}

\def\marksize{1.5};
\def\plotnthpoint{10};
\begin{axis}[
grid=both,
xmin=0.0,
xmax=1,
ymin=0.4,
ymax=1,
xlabel={$I_\mathrm{E,I},I_\mathrm{A,O}$},
ylabel={$I_\mathrm{E,O},I_\mathrm{A,I}$},
width=\linewidth,
height=0.55\linewidth,
]

\addplot[each nth point=\plotnthpoint,only marks,mark size =\marksize pt, color=mittelblau] table [x=x_1, y=y_1, col sep=comma] {tikz/results/ae_128_precursor_small_zp_075/traj_scatter_e1.txt};

\addplot[each nth point=\plotnthpoint,only marks,mark size =\marksize pt, color=rot] table [x=x_2, y=y_2, col sep=comma] {tikz/results/ae_128_precursor_small_zp_075/traj_scatter_e2.txt};

\addplot+ [mittelblau,mark=none] 
table[x=mi ,y=e1,col sep=comma]{./tikz/results/ae_128_precursor_small_zp_075/exit.txt};

\addplot+ [rot,mark=none] 
table[x=e2 ,y=mi,col sep=comma]{./tikz/results/ae_128_precursor_small_zp_075/exit.txt};

\node[rotate=10] at  (axis cs:0.5, 0.82) {\footnotesize $E_\mathrm{b}/N_0=4\,\text{dB}$};

\draw[-latex, line width=1.0pt, color=black] (axis cs:0.65, 0.47) -- node[below,pos=0.1,black]{\footnotesize Outer autoencoder} (axis cs:0.6, 0.54);

\draw[-latex, line width=1.0pt, color=black] (axis cs:0.3, 0.9) -- node[above,pos=0.1,black]{\footnotesize Inner autoencoder} (axis cs:0.35, 0.77);

\end{axis}

\end{tikzpicture}}
	\vspace{-0.1cm}
	\caption{\footnotesize Scattered EXIT chart for ($k=64$, $n=128$). Each point represents an encountered input/output mutual information in a trajectory.
	}
	\label{fig:scattered_exit_chart}
	\vspace{-0.2cm}
\end{figure}
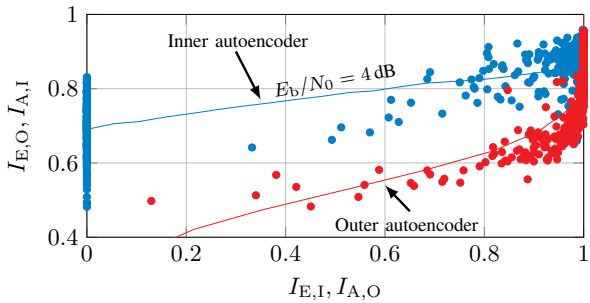

The outer autoencoder is trained as in \cite{cammerer2019tcom}, where it was found that training in the waterfall region leads to the best overall performance. 
This means that we train the outer autoencoder to be as good as possible and then fit the inner autoencoder to the resulting outer autoencoder.
Moreover, the proposed design via \ac{EXIT} charts allows to fit autoencoders for large block lengths.
We can train the \ac{CNN}-based autoencoders on short block lengths with a certain \ac{EXIT} behavior and evaluate the concatenated autoencoders on large block lengths to achieve scaling to large block lengths. 
Similar to \cite{clausius21serialAE} we propose a two step training process.
First, we train the inner autoencoder with high a priori information $I_\mathrm{A}>0.8$. %
The training process is stopped, once the slope of the EXIT characteristic starts increasing (it starts out flat). 
Second, set a certain fraction $\alpha$ of each batch to $I_\mathrm{A}=0$ and continue training. 
As a result, the autoencoder is optimized for high a priori information, which is to be expected from the outer lower rate code, but a reasonably high decoding performance without a priori information to start the iterative decoding. 
We observed that the fraction $\alpha$ directly relates to the slope of the EXIT characteristic. 
A lower $\alpha$ leads a to steeper slope and vice versa. 
This is visualized in Fig.~\ref{fig:exit_chart} for $\alpha=0.075$ and $\alpha=0.085$ at $E_\mathrm{b}/N_0=4\,$dB. 
Further, another inner autoencoder is shown for $\alpha=0.08$ at $E_\mathrm{b}/N_0=2\,$dB.
Lastly, two trajectories are shown for an interleaver length of $n=128$ and $n=2048$.
We can see that the trajectory for $n=128$ does not converge and Fig.~\ref{fig:scattered_exit_chart} reveals that even for $\alpha=0.075$ at $E_\mathrm{b}/N_0=4\,$dB, the trajectory corridor is quite small. 
This means the for $n=128$ the inner and outer autoencoder are fitted nicely.
The other trajectory in Fig.~\ref{fig:exit_chart} for $n=2048$ converges to the intersection point which is just shy of $1.0$, determining the best possible performance at $E_\mathrm{b}/N_0=2\,$dB even for $n \to \infty$.

\subsection{Discussion Component vs Unfolded Training }

A problem with unfolded training is the computational complexity, as
every unfolded iteration runs thorugh a forward and a backward pass.
In contrast, \ac{TGP} is based on a single iteration and, thus, saves computational complexity and time by a factor of the number of iterations $N_\mathrm{it}$ in the decoder. 
While the unfolding increases complexity, it provides the opportunity for the decoder to mitigate short block length effects, i.e., short cycles in the graph and correlated a priori information.
As the \ac{TGP} trained decoder never experiences correlated a priori information during training, it cannot mitigate these effects. 
An advantage of \ac{TGP}, in terms of convergence over training epochs, is that the input to the decoders during training is always of good quality, since the a priori \acp{LLR} are not output of a previous (potentially untrained) decoder. 
Therefore, the \ac{TGP} approach benefits from a fast convergence, especially at the beginning of training.
Lastly, \ac{TGP} does not directly optimize the \ac{BER} as unfolded training does \cite{cammerer2019tcom}.
While the decoders optimize the \ac{BER} in every iteration, the EXIT behavior of the components determine the performance in terms of \ac{BER} and \ac{BLER}.
We observed that inner components with a steeper slope tend to perform better in terms of \ac{BLER} and worse in \ac{BER} than an inner components with a more gentle slope.
The steeper slope of the inner component leads to a later intersection with the outer component but a tighter corridor for the trajectory, as shown in Fig.~\ref{fig:exit_chart}.
Thus, the outer component can output better bit estimates, if the trajectory converges.
However, the trajectory converges less often due to the tighter corridor.

\section{Results}
\label{sec:results}

For the serial TurboAE the encoders are plain \acp{CNN} and the decoder are \acp{DCCNN} with hyperparameters for architecture and training as in Tab.~\ref{tab:hyperparameter}.
For all evaluations we use interleavers according to the LTE standard \cite{3gpp2007standard}.

\begin{table}
	\caption{\footnotesize Hyperparameters for architecture and training the serial TurboAE}
	\label{tab:hyperparameter}
	\centering
	\vspace{-0.1cm}
	\begin{tabular}{l|l}
		Parameter& Value \\
		\hline
		CNN Layers & 5 \\
		CNN $F,K$ & $100,5$ \\
		DCCNN Blocks & $3$\\ 
		Layers per Bl. & $3$\\ 
		1. DCCNN $F_0,F,K$ & $16,12,5$\\ 
		2. DCCNN $F_0,F,K$ & $16,16,5$\\ 
		3. DCCNN $F_0,F,K$ & $16,12,5$\\ 
		$(k,n)$ & $(64,128)$\\
		Padding & Circ. \cite{ye19circularAE}
		
	\end{tabular}
	\begin{tabular}{l|l}
		Parameter& Value \\
		\hline
		Loss & BCE \\
		$T_{\mathrm{TX}}$, $T_{\mathrm{RX}}$  & $100$, $500$  \\
		Batch size & $500-2000$  \\
		Learning rate & $10^{-4}-10^{-6}$  \\
		Encoder SNR & $4.0\,\mathrm{dB}$  \\
		Enc. $I_\mathrm{A}$ & $0.8-1$  \\
		Decoder SNR & $0.5-4.0\,\mathrm{dB}$ \\
		Dec. $I_\mathrm{A}$ & $0.6-0.9$ \\
		$\alpha$ & $0.075$ 
	\end{tabular}
	\vspace{-0.4cm}
\end{table}

\subsection{Decoding Performance for Large Block Lengths}

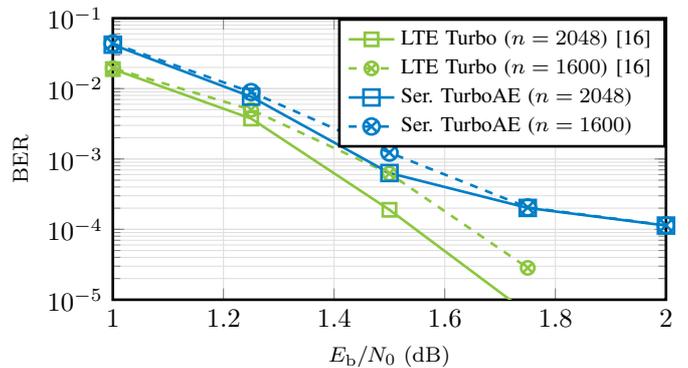
\begin{figure}[t]
	\centering
	\begin{tikzpicture}
		
		\pgfplotsset{compat=1.5}

		\begin{axis}[
			xmode=normal,
			ymode=log,
			xlabel=\footnotesize $E_\mathrm{b}/N_0~(\mathrm{dB})$, %
			ylabel=\footnotesize $\mathrm{BER}$,
			xmin = 1,
			xmax=2.0,
			ymax=10^(-1),
			ymin=10^(-5),
			mark size=2.5pt,
			legend style={at={(axis cs:2,0.1)},anchor=north east}, 
			grid=both,
			minor grid style={gray!25},
			major grid style={gray!25},
			width=\linewidth,
	        height=0.6\linewidth,
			cycle list name=corporate colours markers,
			legend cell align={left},
			line width=1.0pt, %
			]

			\addplot+ [apfelgruen, mark=square, mark options={ solid}]
			table[x=snr,y=ber, ,col sep=comma]{./tikz/results/bler/turbo_n2048_k1024_it8.txt};
			\label{plot:turbo_lte_2048}
            \addlegendentry{\footnotesize LTE Turbo ($n=2048$) \cite{3gpp2007standard}};
            
            \addplot+ [apfelgruen, mark=otimes, dashed,mark options={ solid}]
			table[x=snr,y=ber, ,col sep=comma]{./tikz/results/bler/turbo_n1600_k800_it8.txt};
			\label{plot:turbo_lte_1600}
            \addlegendentry{\footnotesize LTE Turbo ($n=1600$) \cite{3gpp2007standard}};

			\addplot+ [mittelblau, mark options={fill=mittelblau, solid},mark=square, mark size= 3pt] 
			table[x expr=\thisrowno{0} ,y=ber,col sep=comma]{./tikz/results/ae_128_precursor_small_snr2_065_precursor_better_decoder_finetune_zpa_08/results_clip0100_it16_lte_n2048.csv};
			\label{plot:tgp_2048}
            \addlegendentry{\footnotesize Ser. TurboAE ($n=2048$)};
            
            \addplot+ [mittelblau, dashed, mark options={fill=mittelblau, solid},mark=otimes, mark size= 3pt] 
			table[x expr=\thisrowno{0} ,y=ber,col sep=comma]{./tikz/results/ae_128_precursor_small_snr2_065_precursor_better_decoder_finetune_zpa_08/results_clip0100_it16_lte_n1600.csv};
			\label{plot:tgp_2048}
            \addlegendentry{\footnotesize Ser. TurboAE ($n=1600$)};

		\end{axis}

	\end{tikzpicture}
    \vspace{-0.35cm}
	\caption{\footnotesize BER over an AWGN channel for long codes ($R=0.5$), based on short ($n=128$) component codes.}
	\label{fig:bler_long}
 	\vspace{-0.3cm}
\end{figure}

Fig.~\ref{fig:bler_long} shows the \ac{BER} of the serial TurboAE with \acf{TGP} compared to the LTE turbo code for large block lengths ($k\in\{800,\,1024\}$, $n\in\{1600,\,2048\}$, $R=\nicefrac{1}{2}$).
To the best of our knowledge, this is the first time that a competitive performance of an autoencoder is demonstrated  for a message length of $k \approx 1000$.
The performance for the serial TurboAE is evaluated with the outer autoencoder and inner autoencoder ($E_\mathrm{b}/N_0=2\,\text{dB}$,~$\alpha=0.08$) shown in Fig.~\ref{fig:exit_chart}.
For the evaluation we increased the number of inputs to the \acp{CNN} and increased the size of the interleaver.
Also the turbo-product framework \cite{turboproduct98pyndiah} can be used in case the \acp{NN} are not based on \acp{CNN}.
Note, the components are trained with $k=64,~n=128$ and are not trained or finetuned for longer lengths.
While the performance is roughly $0.1\,$dB worse than the LTE turbo code in the low \ac{SNR} regime, the serial TurboAE shows an error floor for higher \acp{SNR}. 
This error floor behavior is due to the intersection in Fig.~\ref{fig:exit_chart} being just shy of a mutual information of $1.0$.

\subsection{Comparison of TGP and Unfolded Training}

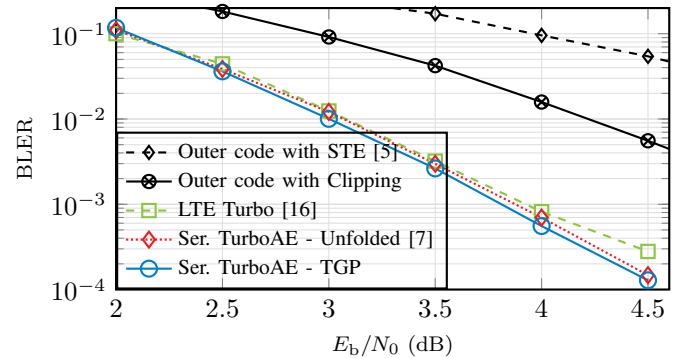
\begin{figure}[t]
	\centering
	\begin{tikzpicture}
		
		\pgfplotsset{compat=1.5}

		\begin{axis}[
			xmode=normal,
			ymode=log,
			xlabel=\footnotesize $E_\mathrm{b}/N_0~(\mathrm{dB})$, %
			ylabel=\footnotesize $\mathrm{BLER}$,
			xmin = 2,
			xmax=4.6,
			ymax=2*10^(-1),
			ymin=10^(-4),
			mark size=2.5pt,
			legend style={at={(axis cs:2,0.0001)},anchor=south west,fill=none}, 
			grid=both,
			minor grid style={gray!25},
			major grid style={gray!25},
			width=\linewidth,
	        height=0.6\linewidth,
			cycle list name=corporate colours markers,
			legend cell align={left},
			line width=0.8pt, %
			]

			\addplot+ [black, mark=diamond, dashed, mark options={ solid}]
			table[x=snr,y=bler, ,col sep=comma]{./tikz/results/bler/ae_outer_ste_more_points.txt};
			\label{plot:ste_binarization}
            \addlegendentry{\footnotesize Outer code with STE \cite{jiang19turboae}};
            
            \addplot+ [black, mark=otimes, mark options={ solid}]
			table[x=snr,y=bler, ,col sep=comma]{./tikz/results/bler/ae_outer_clip_more_points.txt};
			\label{plot:graduell_binarization}
            \addlegendentry{\footnotesize Outer code with Clipping};

				\addplot+ [apfelgruen, mark=square, dashed, mark options={ solid}]
			table[x=snr,y=bler, ,col sep=comma]{./tikz/results/bler/turbo_n64_k128_it6.txt};
			\label{plot:turbo_lte_maxlog_64_128}
            \addlegendentry{\footnotesize LTE Turbo \cite{3gpp2007standard}};

			\addplot+ [rot, densely dotted, mark options={fill=rot, solid},mark=diamond, mark size= 3pt] 
			table[x expr=\thisrowno{0} ,y=bler,col sep=comma]{./tikz/results/bler/stae_result_lte_interleaver.txt};
			\label{plot:uf_n6}
            \addlegendentry{\footnotesize  Ser. TurboAE - Unfolded \cite{clausius21serialAE}};

            \addplot+ [mittelblau, mark options={fill=mittelblau, solid},mark=o, mark size= 3pt] 
			table[x expr=\thisrowno{0} ,y=bler,col sep=comma]{./tikz/results/ae_128_precursor_small_zp_075/results_clip0100_it6_n128_lte.csv};
			\label{plot:tgp_n6}
            \addlegendentry{\footnotesize Ser. TurboAE - TGP };

		\end{axis}

	\end{tikzpicture}
    \vspace{-0.35cm}
	\caption{\footnotesize  BLER over an AWGN channel ($k=64$, $n=128$, $N_\mathrm{it}=6$).}
	\label{fig:bler}
	\vspace{-0.35cm}
\end{figure}

\begin{figure}[t]
	\centering
	\resizebox{\columnwidth}{!}{\begin{tikzpicture}
		
		\pgfplotsset{compat=1.5}

		\begin{axis}[
			xmode=normal,
			ymode=log,
			xlabel=\footnotesize Epochs, %
			ylabel=\footnotesize $\mathrm{BER}$,
			xmin = 0,
			xmax=1200,
			ymax=10^(-2),
			mark size=2.5pt,
			legend style={at={(axis cs:1200,0.01)},anchor=north east}, 
			grid=both,
			minor grid style={gray!25},
			major grid style={gray!25},
			width=\linewidth,
	        height=0.40\linewidth,
			cycle list name=corporate colours markers,
			legend cell align={left},
			line width=0.8pt, %
			each nth point=20,
			]

\addplot+ [apfelgruen, mark=none, densely dotted, mark options={ solid}]
	table[x=x,y=val_ber, ,col sep=comma]{./tikz/results/training/ae_unfolded.txt};
			\label{plot:unfolded_training}
            \addlegendentry{\footnotesize Unfolded AE};
            
\addplot+ [mittelblau, mark=none, dashed, mark options={ solid}]
	table[x=x,y=val_ber, ,col sep=comma]{./tikz/results/training/ae_outer.txt};
			\label{plot:tgp_outer_training}
            \addlegendentry{\footnotesize Outer AE};

\addplot+ [rot, mark=none, solid, mark options={ solid}]
	table[x=x,y=val_ber, ,col sep=comma]{./tikz/results/training/ae_inner.txt};
			\label{plot:tgp_inner_training}
            \addlegendentry{\footnotesize Inner AE};

		\end{axis}

	\end{tikzpicture}}
    \vspace{-0.45cm}
	\caption{\footnotesize Convergence speed comparison in terms of BER vs. training epochs ($k=64$, $n=128$).}
	\label{fig:convergence speed}
	\vspace{-0.45cm}
\end{figure}
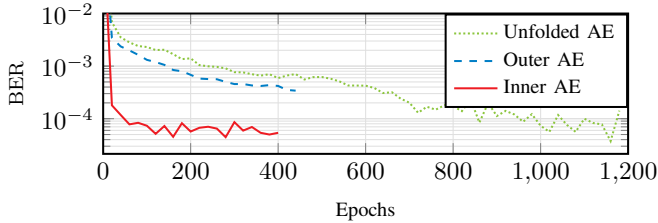

The decoding performance after training is shown in Fig.~\ref{fig:bler} in terms of \ac{BLER} for short block lengths ($k=64$, $n=128$, $R=\nicefrac{1}{2}$).
The serial TurboAE with \ac{TGP} and $N_\mathrm{it}=6$ iterations performs slightly better than the serial TurboAE with unfolded training.
Both perform marginally better than the LTE turbo code.
Further, the outer code with the proposed gradual clipping outperforms the \ac{STE}-based proposition from \cite{jiang19turboae}.
A more prominent advantage shows the training behavior which is displayed in Fig.~\ref{fig:convergence speed}. 
The figure compares the valdiation \ac{BER} of the components with \ac{TGP} and the unfolded training. Note, the \ac{BER} of the outer code is without iterative decoding and the inner code is with iterative decoding.
Both need significantly fewer epochs for training than the unfolded training.
Finally, we observed that component-wise \ac{TGP} is more robust in a sense that the component autoencoders consistently converged to the same shown performance, while for the unfolded training, the shown results are from the best run out of many.

\section{Discussion on Model Complexity}

A valid criticism of autoencoders for channel coding is the increased computational and memory complexity. 
The proposed systems \cite{jiang19turboae, clausius21serialAE, productae22jamali} consists of a huge amount of weights, see Tab.~\ref{tab:num_weights}. 
However, the complexity is simply outside the scope of these works. 
Therefore, these works can be seen as design time systems, where a great exploration space is needed to find good solutions.
This exploration space of a \ac{NN} is usually coupled to the number of trainable weights. 
However, once a good solution is found, the knowledge can be distilled into smaller \acp{NN} \cite{hinton2015distilling}.
The intuition is that the solution is inside the solution space of the larger and the smaller \ac{NN}.
However, the smaller \ac{NN} can not converge to the solution, as the needed convergence trajectory is not inside the solution space. 
A common knowledge distillation approach is the student-teacher approach.
Here, the student is influenced by a regression loss with the outputs of the teacher.
We applied this approach to the encoder with an \ac{MSE} loss.
A smaller \ac{CNN}-based encoder with a total of $148$ weights (reduction by $99.96\,\%$) for the inner and outer encoder in total was able to represent the solution from the large encoder without any loss in performance.

\begin{table}
	\caption{\footnotesize Approximate number of weights}
	\vspace{-0.1cm}
	\label{tab:num_weights}
	\centering
	\begin{tabular}{l|l|l|l}
		Name& Encoder& Decoder & Layer type\\
		\hline
		TurboAE \cite{jiang19turboae}& $3.0\cdot10^4$ & $2.4\cdot10^6$ & Conv.\\
		Ser. TurboAE \cite{clausius21serialAE}& $4.1\cdot10^5$ & $2.4\cdot10^6$& Conv.\\
		ProductAE \cite{productae22jamali} & $2.8\cdot10^5$ & $1.9\cdot10^6$& Dense\\
		\emph{This work} & $148$ & $1.2\cdot10^5$& Conv.
		\vspace{-0.25cm}
	\end{tabular}
\vspace{-0.2cm}
\end{table}

\section{Conclusion}
\label{sec:conclusion}
We introduced the component-wise \ac{TGP} for the serial TurboAE and demonstrated a faster and more consistent training. 
By training the components to match a desired \ac{EXIT} curve, the resulting serial TurboAE shows a competitive \ac{BER} for message lengths of $k\approx 1000$.
To the best of our knowledge, this is the first autoencoder that performs close to classical codes in this regime.
Furthermore, we proposed a new quantization strategy to force a trainable encoder to a \ac{BPSK} modulation.
Lastly, we demonstrated a reduction of the number of weights by $99.96\,\%$ in the encoder after training.
Future works could include the exploration of different autoencoders as components and complexity reduction in the decoder.

\bibliographystyle{IEEEtran}
\bibliography{IEEEabrv,references}

\end{document}